\documentclass{ws-procs975x65}

\usepackage{amsmath}
\usepackage{amsfonts}

\newcommand{\be}{\begin{equation}}
\newcommand{\ee}{\end{equation}}

\newcommand{\pa}{\partial}
\newcommand{\ve}{\varepsilon}
\newcommand{\lb}{\boldsymbol{(}}
\newcommand{\rb}{\boldsymbol{)}}

\def\omk{\omega_{\text{kinetic}}}
\def\oms{\omega_{\text{static}}}

\def\hn{H_{\text{N}}}
\def\hi{H_{\text{1PN}}}
\def\hii{H_{\text{2PN}}}
\def\hiii{H_{\text{3PN}}}
\def\hiiireg{H_{\text{3PN}}^{\text{Had}}}

\def\htt#1#2{h^{\rm TT}_{#1#2}}
\def\pitt#1#2{\pi_{\mathrm{TT}}^{#1#2}}
\def\piti#1#2{\widetilde{\pi}^{#1#2}}

\begin{document}

\title{DIMENSIONAL REGULARIZATION OF THE GRAVITATIONAL INTERACTION OF POINT MASSES
IN THE ADM FORMALISM\footnote{The research of P.J.\ has been partially supported by the KBN Grant no 1 P03B 029 27.}}

\author{Thibault Damour}
\address{
Institut des Hautes \'Etudes Scientifiques,
Bures-sur-Yvette, France\\
E-mail: damour@ihes.fr}

\author{Piotr Jaranowski}
\address{Institute of Theoretical Physics,
University of Bia{\l}ystok,
Bia{\l}ystok, Poland\\
E-mail: pio@alpha.uwb.edu.pl}

\author{Gerhard Sch\"afer}
\address{
Theoretisch-Physikalisches Institut,
Friedrich-Schiller-Universit\"at,
Jena, Germany\\
E-mail: Gerhard.Schaefer@uni-jena.de}

\begin{abstract}

The ADM formalism for two-point-mass systems in $d$ space dimensions is sketched. It is pointed out that the regularization ambiguities of the 3rd post-Newtonian ADM Hamiltonian considered directly in $d=3$ space dimensions can be cured by dimensional continuation (to complex $d$'s), which leads to a finite and unique Hamiltonian as $d\to3$. Some so far unpublished details of the dimensional-continuation computation of the 3rd post-Newtonian two-point-mass ADM Hamiltonian are presented.

\end{abstract}

\keywords{binary systems, equations of motion, point masses, dimensional regularization}

\bodymatter

\section{Introduction}

The problem of finding the equations of motion (EOM) of a two-body system within the post-Newtonian (PN) approximation of general relativity is solved up to the 3.5PN order of approximation for the case of compact and nonrotating bodies [by $n$PN approximation we mean corrections of order $(v/c)^{2n}\sim\lb{Gm/(rc^2)}\rb^n$ to Newtonian gravity]. The 3PN level of accuracy was achieved only recently. There exist two independent derivations of the 3PN EOM using distributional (Dirac delta's) sources: either ADM-Hamiltonian-based\cite{DJS,DJS01a}, or harmonic-coordinate-based.\cite{BF,BDEF04} There also exists a third independent derivation of the 3PN EOM in harmonic coordinates using a surface-integral approach.\cite{IF}

To cure the self-field divergencies of point particles it is necessary to use some regularization method. It turned out that different such methods applied in $d=3$ space dimensions were not able to give unique EOM at the 3PN order. Only by employing dimensional continuation was it possible to obtain unambiguous results.\cite{DJS01a,BDEF04} In this note we review the dimensional-continuation-based derivation of the 3PN two-point-mass ADM Hamiltonian.

\section{ADM formalism for 2-point-mass systems in $d$ space dimensions}

We use units such that $c{=}16\pi{G_{d+1}}{=}1$. We work in an asymptotically flat $(d+1)$-dimensional spacetime with Minkowskian coordinates $x^0$, ${\bf x}{\equiv}(x^1,\ldots,x^d)$. Particles are labeled by the index $a\in\{1,2\}$; masses, positions, and momenta of the particles are denoted by $m_a$, ${\bf{x}}_a{\equiv}(x_a^1,\ldots,x_a^d)$, and ${\bf{p}}_a{\equiv}(p_{a1},\ldots,p_{ad})$, respectively. We also define: ${\bf r}_a:={\bf x}-{\bf x}_a$, $r_a:={\vert{\bf r}_a\vert}$, ${\bf n}_a:={\bf r}_a/r_a$; ${\bf{r}}_{12}:={\bf{x}}_1-{\bf{x}}_2$, $r_{12}:=\vert{\bf{r}}_{12}\vert$ ($\vert{\bf{v}}\vert$ means here the Euclidean length of the $d$-vector ${\bf{v}}$).
The canonical variables of the theory consist of {\em matter} variables $({\bf x}_a,{\bf p}_a)$ and {\em field} variables $(\gamma_{ij},\pi^{ij})$, where the space metric $\gamma_{ij}$ is induced by the full space-time metric on the hypersurface $x^0$=const; its conjugate $\pi^{ij}$ can be expressed in terms of the extrinsic curvature of that hypersurface.

Source terms in the {\em constraint equations} written down for two-point-mass systems are proportional to the $d$-dimensional Dirac delta functions $\delta({\bf x}-{\bf x}_a)$. We use the ADM gauge defined by the conditions (TT $\equiv$ transverse-traceless):
\be
\gamma_{ij} = \bigg(1+\frac{d-2}{4(d-1)}\phi\bigg)^{4/(d-2)}\delta_{ij} + {\htt ij},
\quad
\pi^{ii} = 0.
\ee
The field momentum $\pi^{ij}$ splits into a TT part ${\pitt ij}$ and a rest ${\piti ij}$ (traceless but expressible in terms of a vector), $\pi^{ij} = {\piti ij} + {\pitt ij}$. If both the constraint equations and the gauge conditions are satisfied, the ADM Hamiltonian can be put into its {\em reduced} form:
\be
H\big({\bf x}_a,{\bf p}_{a},{\htt ij},{\pitt ij}\big)
= -\int\mathrm{d}^dx \, \Delta\phi\big({\bf x}_a,{\bf p}_{a},{\htt ij},{\pitt ij}\big).
\ee
The PN expansion of the reduced Hamiltonian is worked out up to the 3.5PN order:
\be
H = \sum_{a=1}^2 m_a + \hn + \hi + \hii + H_{\text{2.5PN}} + \hiii + H_{\text{3.5PN}} + {\cal O}\lb(v/c)^8\rb.
\ee

\section{Dimensional regularization of the 3PN Hamiltonian}

In Refs.\ \refcite{DJS} it was shown that the Riesz-implemented Hadamard regularization of the 3PN two-point-mass Hamiltonian performed in $d=3$ space dimensions gives ambiguous results. The ambiguities were parametrized by two numerical coefficients called ambiguity parameters and denoted by $\omk$ and $\oms$.

Dimensional continuation consists in obtaining the 3-dimensional Hamiltonian as $\lim_{d\to3}\hiii(d)$, where $\hiii(d)$ is the Hamiltonian computed in $d$ space dimensions. This can be done straightforwardly if no poles proportional to $1/(d-3)$ arise when $d\to3$ (or if one shows that these poles can be renormalized away, as happens in harmonic coordinates\cite{BDEF04}).
Reference \refcite{DJS01a} has shown that out of all terms building up the Hamiltonian density there are ten terms $T_A(d)$, $A=1,\ldots,10$, giving rise to poles when $d\to3$. It was checked that the poles produced by these terms cancel each other, thus $\lim_{d\to3}\hiii(d)$ exists. Moreover, it was shown that for all other terms the 3-dimensional regularization give the same results as dimensional continuation.

Let $\hiiireg$ be the 3PN Hamiltonian obtained in Refs.\ \refcite{DJS} by using an Hadamard ``partie finie'' (Pf) regularization defined in $d=3$ space dimensions. To correct this Hamiltonian one needs to compute the difference $\Delta{\hiii}:=\lim_{d\to3}\,\hiii(d)-\hiiireg$. Only ten terms $T_A$ contribute to $\Delta{\hiii}$, therefore
\be
\label{dh}
\Delta{\hiii} = \lim_{d\to3}\,\int\!\!{\mathrm{d}}^dx \sum_{A=1}^{10} T_A(d)
- \text{Pf}\int\!\!{\mathrm{d}}^3x \sum_{A=1}^{10} T_A(3).
\ee
Below we present three different methods which we used to compute $\Delta{\hiii}$. The details of the 2nd and 3rd method were not published so far.
Knowing $\Delta{\hiii}$ one determines the values of both ambiguity parameters: $\omk=41/24$, $\oms=0$.

{\bf 1st method.} In Ref.\ \refcite{DJS01a} $\Delta{\hiii}$ was computed by means of the analysis of the local behaviour of the terms $T_A$ around the particle positions ${\bf{x}}={\bf{x}}_a$.

{\bf 2nd method.} It is possible to compute all $d$-dimensional integrals in Eq.\ \eqref{dh} explicitly. To do this one uses the Riesz formula
\be
\label{riesz}
\int\mathrm{d}^dx \: r_1^\alpha\,r_2^\beta = \pi^{d/2}
\frac{\Gamma\lb(\alpha+d)/2\rb \Gamma\lb(\beta+d)/2\rb \Gamma\lb-(\alpha+\beta+d)/2\rb}
     {\Gamma(-\alpha/2) \Gamma(-\beta/2) \Gamma\lb(\alpha+\beta+2d)/2\rb} r_{12}^{\alpha+\beta+d},
\ee
and the distributional differentiation of homogeneous functions, e.g.,
\be
\frac{\pa^2}{\pa{x^i}{x^j}} \frac{1}{r_a^{d-2}}
= \text{Pf} \Big( (d-2)\frac{d\,n_a^i n_a^j-\delta_{ij}}{r_a^d} \Big)
- \frac{4\pi^{d/2}}{d\,\Gamma(d/2-1)} \delta_{ij} \delta({\bf x}-{\bf x}_a).
\ee

{\bf 3rd method.} Instead of $d$-dimensional Dirac distributions $\delta$ one uses $d$-dimensional Riesz kernels $\delta_{\ve_a}$ to model point particles:
\be
\delta({\bf x}-{\bf x}_a) = \lim_{\ve_a\to0} \delta_{\ve_a}({\bf x}-{\bf x}_a),
\quad
\delta_{\ve_a}({\bf x}-{\bf x}_a) := \frac{\Gamma\lb(d-\ve_a)/2\rb}{\pi^{d/2}\,2^{\ve_a}\,\Gamma(\ve_a/2)} r_a^{\ve_a-d}.
\ee
Then one uses the formula \eqref{riesz} to calculate the integrals in Eq.\ \eqref{dh} and, at the end of the calculation, one takes the limit $\ve_1\to0$, $\ve_2\to0$. No distributional differentiation is needed.

We have shown that these three methods yield the same final results.


\begin{thebibliography}{9}

\bibitem{DJS}
P.\ Jaranowski and G.\ Sch\"afer,
{\em Phys.\ Rev.\ D} {\bf 57}, 7274 (1998); {\bf 63}, 029902(E) (2001);
{\bf 60}, 124003 (1999);
T.\ Damour, P.\ Jaranowski, and G.\ Sch\"afer,
{\em{ibid.}} {\bf 62}, 044024 (2000);
{\bf 62}, 021501(R) (2000); {\bf 63}, 029903(E) (2001).

\bibitem{DJS01a}
T.\ Damour, P.\ Jaranowski, and G.\ Sch\"afer,
{\em Phys.\ Lett.\ B} {\bf 513}, 147 (2001).

\bibitem{BF}
L.\ Blanchet and G.\ Faye,
{\em Phys.\ Lett.\ A} {\bf 271}, 58 (2000);
{\em Phys.\ Rev.\ D} {\bf 63}, 062005 (2001).

\bibitem{BDEF04}
L.\ Blanchet, T.\ Damour, and G.\ Esposito-Far\`ese,
{\em Phys.\ Rev.\ D} {\bf 69}, 124007 (2004).

\bibitem{IF}
Y.\ Itoh and T.\ Futamase,
{\em Phys.\ Rev.\ D} {\bf 68}, 121501(R) (2003);
Y.\ Itoh,
{\em{ibid.}} {\bf 69}, 064018 (2004).

\end{thebibliography}
\end{document}